\documentclass[aps,prb,reprint,longbibliography,superscriptaddress]{revtex4-2}

\usepackage[colorlinks=true,citecolor=blue]{hyperref}
\usepackage{amsmath,latexsym,amssymb,bm,color,graphicx,multirow,array,wrapfig,caption,scalerel,subcaption}

\graphicspath{{figures/}}

\DeclareMathOperator{\tr}{tr}
\begin{document}

\title{Strange Correlator for 1D Fermionic Symmetry-Protected Topological Phases}

\author{Yinning Niu}
\affiliation{State Key Laboratory of Surface Physics, Fudan University, Shanghai 200433, China}
\affiliation{Center for Field Theory and Particle Physics, Department of Physics, Fudan University, Shanghai 200433, China}

\author{Yang Qi}
\affiliation{State Key Laboratory of Surface Physics, Fudan University, Shanghai 200433, China}
\affiliation{Center for Field Theory and Particle Physics, Department of Physics, Fudan University, Shanghai 200433, China}

\begin{abstract}
    Strange correlators are useful tools for diagnosing symmetry-protected topological states from their bulk wave functions.
    We study strange correlators for one-dimensional fermionic symmetry-protected topological states using fixed-point wave functions, and show that a combination of strange correlators constructed using fermion annihilation operators and bosonic order parameters can fully diagnose the classification from such wave functions.
    By converting the strange correlator to a correlation in a one-dimensional statistical problem described by transfer matrices, we show that when the wave function is topologically nontrivial, the corresponding strange correlator exhibits a long-range-order behavior, which can be analyzed from the symmetry properties of the transfer matrices constructed from the fixed-point wave function.
    Our general argument is demonstrated by several concrete examples.
\end{abstract}

\maketitle

\section{Introduction}
Symmetry-protected topological (SPT) phases~\cite{ChenScience2012,chen_symmetry_2013}, including the topological insulators and topological superconductors~\cite{HasanRMP2010, XLQiRMP2011},
are short-range-entangled topological phases~\cite{XChenLUT2010} beyond Landau's paradigm.
An important feature of SPT phases is the bulk-boundary correspondence:
when the bulk is topologically nontrivial, its boundary will exhibit anomalous behaviors, including symmetry-protected gaplessness.
(The boundary can also exhibit symmetry breaking or realize a gapped topological order with an anomalous symmetry enrichment in 2+1 dimensions.)
Therefore, a gapless surface is usually used as a signature to detect a topologically nontrivial bulk.
However, this signature is sometimes inconvenient or impossible to use:
In numerical simulations such as exact diagonalization and quantum Monte Carlo, the extra computational cost of simulating a boundary (instead of the common practice of using periodic boundary conditions) is often unacceptable;
it is impossible to design fully symmetric boundaries for SPT phases protected by crystalline~\cite{FuTCI2011} or average symmetries~\cite{RMaASPT2023}.
Therefore, it is desirable to have bulk-detection tools for nontrivial SPT phases.

Unlike free-fermion topological states, which can be detected using topological invariant~\cite{ChiuRMP2016} (such as the Chern number and the Fu-Kane invariant~\cite{LFuPRB2006}) computed from the bulk band structure, the bulk detection of interacting SPT states is more challenging.
In principle, SPT phases can be detected by evaluating the partition function on nontrivial space-time manifold with symmetry fluxes~\cite{DijkgraafWitten, witten_fermion_2016}, or equivalently by gauging the symmetry and study the resulting topological order~\cite{levin_braiding_2012,CJWangPRL2014,MChengPRX2018}.
However, this tool could be inconvenient to use in practice, due to the difficulties of generalzing a lattice model to curved manifolds, and evaluating the partition function for a realistic model.
In one dimension, SPT states can be detected using symmetry-transformation properties of the Schmidt eigenstates of the entanglement Hamiltonian~\cite{Pollmann2010Entangle}, but this approach is difficult to generalize to higher dimensions.
Another useful tool for bulk detection is the strange correlator first proposed by \citet{you_wave_2014}, which will be the main focus of this work.
Defined as a correlation function evaluated between two different SPT wave functions, it has proven to be a powerful bulk diagnostic tool for identifying nontrivial topological properties of symmetry protected topological phases~\cite{HQWu2015,Vanhove2018,CKZhou2022,Lepori2023,zhang2022strange,lee2023symmetry}, because it uses the ground state wave function which is easily obtained in numerical simulations, and it can be generalized to higher dimensions.


A strange correlator is generally defined as the correlation function of two local observable operators $\hat O(r)$ and $\hat O(r')$, inserted \begin{equation}
\label{eq:strange-corr}
C(r,r') = \frac{\langle \Omega | \hat O(r) \hat O(r') | \Psi \rangle}{\langle \Omega | \Psi \rangle}
\end{equation}
between two different states $|\Phi\rangle$ and $|\Psi\rangle$.
$|\Psi\rangle$ usually is chosen to be the SPT wave function to be diagnosed, and $|\Phi\rangle$, dubbed the reference state, can be the trivial SPT state or another SPT state whose classification is known.
The strange correlator then measures the difference between the two states:
If both states belong to the same SPT phase, the strange correlator decay exponentially as $|r - r'|\rightarrow\infty$; if they belong to different SPT phases, one can design suitable observable such that the strange correlator does not decay exponentially: it either goes to a constant or decays as a power-law of $|r-r'|$, which we generally refer to as long-range-order behavior in this work~\footnote{In fact, the power-law behavior is better addressed as a quasi-long-range order.
However, since this work focuses on 1D systems, where the power-law behavior is less common than in higher-dimensions, we will address them simply as long-range order.}.
Intuitively, the strange correlator can be viewed as a correlation function on a ``temporal'' boundary between the two states of $|\Phi\rangle$ and $|\Psi\rangle$, and therefore is able to detect the gapless nature of the boundary state if the two states belong to different topological phases.
Hence, the strange correlator not only serves as a useful diagonostic tool for detecting SPT phases, especially in situatios where other approaches are not convenient, but also provides interesting connection between wave functions of topological phases and correlation functions in anomalous gapless effective theories such as conformal field theories~\cite{Scaffidi2016}.
However, previous works on the strange correlator usually focus specific examples, and a general proof or comprehensive theoretical framework for the strange correlator method is still lacking, especially for fermionic SPT states.


In this work, we study the strange correlator in one-dimensional (1D) fermionic SPT (FSPT) phases~\cite{gu_tensor-entanglement-filtering_2009,chen_classification_2011,chen_complete_2011,gu_symmetry-protected_2014,wang_towards_2018,wang_construction_2020}.
Using relations between 1D strange correlators and correlation function in 1D statistical systems,
we give a sysmetical understanding of long-range behaviors of strange correlators.
We observe that, the strange correlator can detect nontrivial bosonic 1D SPT states because the corresponding statistical system exhibits degeneracy due to the nontrivial projective representation associated with the nontrivial bosonic phase factors. 
Furthermore, we show that a strange correlator, constructed using fermion operators, can detect nontrivial fermionic decoration in an FSPT state.
These results confirms the expectation to the long-range behavior of strange correlator in Ref.~\cite{you_wave_2014}.

The rest of this paper is organized as follows:
We first study the simple case where the fermion symmetry group $G_f$ is a direct product of bosonic symmetry group $G_b$ and the fermion-parity symmetry group $\mathbb Z_2^f$ in Sec.~\ref{sec:gb-z2f}.
In Sec.~\ref{sec:wavefunction}, we briefly review the fixed-point wave function for FSPT states.
In Sec.~\ref{sec:bosonic}, we study the strange correlator in 1D bosonic SPT states.
In particular, we develop a framework of relating the strange correlator to symmetry properties of transfer matrices determined from the fixed-point wave function, which will be used in our analysis of fermionic strange correlators.
In Sec.~\ref{sec:corr-f}, we study the fermionic strange correlator and argue that it can detect whether an FSPT state has a nontrivial fermion decoration.
Finally in Sec.~\ref{sec:diagnose}, we combine the two types of strange correlators, and show how they can be used to fully diagnose an FSPT state.
Next, in Sec.~\ref{sec:gb-z2f-ext}, we generalize the above discussion to the more general cases where $G_f$ is a nontrivial extension of $G_b$ over $\mathbb Z_2^f$, $G_f=G_b\times_{\omega_2}\mathbb Z_2^f$, where $\omega_2\in H^2(G_b,\mathbb Z_2^f)$ is a 2-cocycle classifying such extensions.
In Sec.~\ref{sec:examples}, we study three concrete examples to demonstrate the use of strange correlators.
Finally, conclusion and discussions are given in Sec.~\ref{sec:conclusion}.

\section{1D FSPT states with $G_f=G_b\times \mathbb Z_2^f$}
\label{sec:gb-z2f}

In this section, we discuss strange correlators in 1D FSPT states where the symmetry group $G_f$ is a direct product of bosonic symmetry group $G_b$ and the fermion-parity symmety $\mathbb Z_2^f$.
The results we get will be generalized to cases where $G_f$ is a nontrivial extension of $G_b$ over $\mathbb Z_2^f$ in the next section.

\subsection{Fixed-Point Wavefunction for FSPT Phases in 1D}
\label{sec:wavefunction}

We begin by briefly reviewing the fixed-point wave function for a 1D FSPT state, which will be used later to compute the strange correlators.
The general form of such wave function is given as a superposition over states labeled by group-element and complex-fermion decorations~\cite{wang_construction_2020}:
\newcommand{\wfconfoneD}{
\includegraphics[scale=1.3]{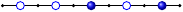}
}
\begin{equation}
\label{eq:1eq2}
\begin{split}
    |\Psi\rangle = \sum_{\text{all conf.}} \Psi\left(
\vcenter{\hbox{\wfconfoneD}}
\right) \\
\stretchleftright{\Bigg|}{\ 
\vcenter{\hbox{\wfconfoneD}}
\ }{\Big\rangle}.
\end{split}
\end{equation}
Here, we first explain the set of states the wave function is built upon, then give the explicit expression of the superposition coefficients.
In the 1D lattice, each vertex $i$ (represented by a small black dot in the above equation) has a bosonic degree of freedom labeled by a group element $g_i \in G_b$.
Each group-element configuration $\{g_i\}$ corresponds to a bosonic basis state $|\{g_i\}\rangle$.
We denote such bosonic basis states by $|\{g_i\}\rangle$.
The center of each bond between nearest-neignbor vertices $i$ and $i+1$ has a complex-fermion degree of freedom denoted by $c_{i,i+1}$ (represented by a big blue dot in the above equation).
The Hilbert space we consider is a tensor product between the Fock space of these fermion modes, and the bosonic Hilbert space spanned by the basis states labeled by group-element configurations.
In other words, the basis states are labeled by a fermion occupation number $n_{i,i+1}$ on each bond $\langle i,i+1\rangle$, and a group element $g_i$ on each vertex, which we denote by $|\{n_{i,i+1}\};\{g_i\}\rangle$.
Using the fermion-creation operator, such a Fock state is expressed as
\begin{equation}
    \label{eq:cgi}
    |\{n_{i,i+1}\};\{g_i\}\rangle
    =\prod_i \left(c_{i,i+1}^\dagger\right)^{n_{i,i+1}}
    |\{g_i\}\rangle.
\end{equation}

In the superpositionin \eqref{eq:1eq2}, the fermion decoration is determined by the bosonic group-element configuration, as
\begin{equation}
    \label{eq:n1}
    n_{i, i+1}=n_1(g_i, g_{i+1}).
\end{equation}
In this paper, we choose to use homogeneous cochain functions to represent the cochains.

Moreover, the phase factor in the superposition is also determined from the group-element decoration as the following,
\begin{equation}
    \label{eq:1D-phase-factor}
    \Psi(\vcenter{\hbox{\wfconfoneD}})
    =\prod_i \nu_2(e, g_i, g_{i+1}),
\end{equation}
where $\nu_2\in H^2[G_b,\mathrm U(1)]$ is a 2-cocycle, and $e\in G_b$ is the identity element.
Therefore, a fixed-point wave function for a FSPT state labeled by $n_1$ and $\nu_2$ can be expressed as:
\begin{align}
    \label{eq:1Dwf}
    |\Psi\rangle
    =\sum_{\{g_i\}}
    \prod_i \nu_2(e, g_i, g_{i+1})
    \left(c_{i,i+1}^\dagger\right)^{n_1(g_i, g_{i+1})}
    |\{g_i\}\rangle.
\end{align}
In fact, 1D FSPT states with $G_f=G_b\times \mathbb Z_2^f$ are classified by two pieces of data: a 1-cocycle $n_1\in H^1(G_b, \mathbb Z_2)$ specifying the fermion decoration, and a 2-cocycle $\nu_2\in H^2[G_b, \mathrm U(1)]$ representing the bosonic phase factor.

\subsection{Bosonic Strange Correlator}
\label{sec:bosonic}

As a warm-up, we study the strange correlator of bosonic order parameters, for 1D bosonic SPT wave functions.
Not only is this result useful for detecting the bosonic layer ($\nu_2$) of an FSPT state (see Sec.~??), but the notations introduced here will also be used to compute the fermionic strange correlator in the next section.

We consider the following bosonic strange correlator,
\begin{equation}
    \label{eq:bs-corr}
    C(i, j)=\frac{\langle\Omega|\hat O_i\hat O_j|\Psi\rangle}{\langle\Omega\Psi\rangle},
\end{equation}
where $|\Omega\rangle$ is the trivial state ($n_1=0$ and $\nu_2=1$), $|\Psi\rangle$ is a bosonic SPT state with $n_1=0$.
For simplicity, we assume that the order parameters $\hat O_{i,j}$ are diagonal in the group-element basis,
\begin{equation}
    \label{eq:Oi}
    \hat O_i = \sum_{g_i\in G_b}O(g_i)|g_i\rangle\langle g_i|.
\end{equation}
(Correlation functions of diagonal operators can be easily mapped to correlation functions of a classicial statistical problem, as we will show below.)
Using the explicit form of fixed-point wave function in Eq.~\eqref{eq:1Dwf} and assuming the fermion decoration $n_1$ is always zero, we can express the bosonic strange correlator \eqref{eq:bs-corr} as follows,
\begin{equation}
    \label{eq:bs-corr-sum}
    C(i, j)=\frac{\sum_{\{g_i\}}O(g_i)O(g_j)\prod_i\nu_2(e, g_i, g_{i+1})}
    {\sum_{\{g_i\}}\prod_i\nu_2(e, g_i, g_{i+1})}.
\end{equation}
This has the form of a correlation function of a classical system, where $\prod_i\nu_2(e, g_i, g_{i+1})$ is the statistical weight, and $O(g_{i,j})$ is the observables.
This motivates us to rewrite the 2-cocyle $\nu_2(e, g_i, g_{i+1})$ as a transfer matrix, which is widely used to solve 1D statistical models.
We define the transfer matrix $T$ as follows,
\begin{equation}
    \label{eq:T-def}
    T_{ij} = \nu_2(e, g_i, g_{i+1}),
\end{equation}
and write the observable as a diagonal matrix $O_{ij}=\delta_{ij}O(g_i)$.
Eq.~\eqref{eq:bs-corr-sum} is then expressed as
\begin{equation}
    \label{eq:bs-corr-T}
    C(i, j)=\frac{\tr\left(OT^{|i-j|}OT^{N-|i-j|}\right)}
    {\tr\left(T^N\right)}.
\end{equation}

In 1D statistical models, the partition function $Z=\tr(T^N)$ is dominated by the leading eigenvalue of the transfer matrix $T$.
Recall that in 1D statistical models, the eigenvectors and their degeneracy is generally determined by the symmetry property of $T$.
This prompts us to study how $T$ transforms under the symmetry group $G_b$.
In fact, since the fixed-point wave functions is invariant under the action of $G_b$, the partition function $Z=\tr(T^N)$ is also invariant.
However, this does not imply that the transfer matrix $T$ is invariant:
In fact, as we shall see later, for a generic $\nu_2$, $T$ is only invariant under a projective action of $G_b$, and the eigenvectors form projective representation of $G_b$.
Only when $\nu_2$ is trivial does the eigenvectors form linear representation of $G_b$.

First, consider the simple case where $G_b$ is unitary.
We define the following representation matrices $\tilde R(g)$ for group elements $g\in G_b$:
\begin{equation}
    \label{eq:tRg-def}
    \tilde R(g)_{ij} = R(g)_{ij}\nu_2(e, g, g_i)^{-1},
\end{equation}
where $R(g)$ denotes the regular representation of $G_b$,
\begin{equation}
    \label{eq:Rg-def}
    R(g)_{ij} = \delta_{g_i, gg_j}.
\end{equation}
Since $\tilde R(g)$ differs from $R(g)$ by a phase factor, it satisfies the twisted relation
\begin{equation}
    \label{eq:tRg-mul}
    \tilde R(g)\tilde R(h) = \nu_2(e, g, gh)^{-1}\tilde R(gh).
\end{equation}
(The proof of the above equation uses the cocycle condition of $\nu_2$.)
This shows that $\tilde R(g)$ is actually a projective representation of $G_b$ associated with the 2-cocycle $\nu_2^{-1}$.
In fact, as explained in Appendix~\ref{app:corep}, $\tilde R(g)$ forms a twisted group algebra $\mathbb C[G_b]^{\nu_2^{-1}}$, which can be decomposed into irreducible projective representations of $G_b$, just like the regular group algebra $\mathbb C[G_b]$ formed by $R(g)$ can be decomposed into irreducible (linear) representations of $G_b$.

Using the definition of $\tilde R(g)$, the intuition that the partition function is invariant under $G_b$ can be expressed by the following property of the transfer matrix $T$:
\begin{equation}
    \label{eq:TR=RT}
    \tilde R(g) T =T\tilde R(g).
\end{equation}
(The proof of this equation also uses the cocycle condition of $\nu_2$.)
In other words, $T$ is a intertwiner of $\tilde R(g)$.
According to the Schur's lemma, $T$ must be a scalar on (the subspace of ) each irreducible representation of $\tilde R(g)$.
Therefore, each nondegenerate eigenvector of $T$ corresponds to a one-dimensional irreducible representation of $\tilde R(g)$, and each subspace of degenerate eigenvectors of $T$ corresponds to a multi-dimensional irreducible representation of $\tilde R(g)$.

Recall that in a 1D statistical system, the degeneracy of the leading eigenvector (the eigenvector whose eigenvaule has the largest magnitude) determines hoe the correlation function of the symmetry order parameter decays at long distances.
In the denominator, the leading eigenvector(s) with dominate the partition function, such that $\tr(T^N)\sim \lambda_1^N$, where $\lambda_1$ is the eigenvalue of the leading eigenvector.
In the numerator, if the leading eigenvector is nondegenerate (belongs to a 1D irreducible represention of the symmetry group),
the order parameter $O$, with carries a nontrivial irreducible representation, will map the leading eigenvector to a different 1D irreducible representation, which would have an eigenvalue $\lambda_2$ of a smaller magnitude: $|\lambda_2|<|\lambda_1|$.
Hence, the numerator behaves as $\tr(OT^{|i-j}OT^{N-|i-j|})\sim \lambda_1^{N-|i-j|}\lambda_2^{|i-j|}$.
Therefore, the strange correlator should decay exponentially as the following,
\begin{equation}
    \label{eq:Bij-l12}
    C(i, j)\sim \left(\frac{\lambda_2}{\lambda_1}\right)^{|i-j|}.
\end{equation}
On the other hand, if the leading eigenvector is degenerate (belongs to a multi-dimensional irreducible representation), the order parameter $O$ can generally map one vector to another in the same irreducible representation.
Therefore, the strange correlator would converges to a constant,
\begin{equation}
    \label{eq:Bij-c}
    C(i, j)\sim \text{Const}.
\end{equation}

When $\nu_2$ belongs to a nontrivial cohomology class in $H^2[G_b, \mathrm U(1)]$, $\tilde R$ carries projective representations of $G_b$, and the irreducible projective representations are all multi-dimensional, because one-dimensional matrices cannot produce the factors in the twisted relation \eqref{eq:tRg-mul}.
Consequently, the strange correlator exhibits the behavior in Eq.~\eqref{eq:Bij-c}, if an appropriate operator $O$ is chosen.
Here, $O$ needs to act nontrivially within the subspace of the projective representation.
For example, if $G_b=G_{b1}\times G_{b2}$, and $\nu_2$ is a nontrivial 2-cocycle when constrainted on $G_{b1}$ but trivial on $G_{b2}$, then $O$ needs to act nontrivially on $G_{b1}$ and trivially on $G_{b2}$ for the strange correlator to exhibit the behavior in Eq.~\eqref{eq:Bij-c}.
In practice, different possible choices of $O$ (carrying different representations of $G_b$) needs to be tried to fully diagnose whether $\nu_2$ is nontrivial.
(We notice that similar procedures of trying different order parameters for a large symmetry group are need for detecting symmetry breaking in classical statistical systems.)

On the other hand, when $\nu_2$ belongs to the trivial class in $H^2[G_b, \mathrm U(1)]$ (i.e. it is a trivial coboundary), the group algebra $\mathbb C[G_b]^{\nu_2^{-1}}$ is actually isomorphic to the untwisted $\mathbb C[G_b]$, and the representation in $\tilde R$ are isomorphic to a linear representation of $G_b$ after a basis transformation.
Therefore, $\tilde R$ can be decomposed into irreducible linear representations of $G_b$, many of which are one-dimensional.
In fact, if $G_b$ is Abelian, then all of its irreducible representations are one-dimensional; when $G_b$ is non-Abelian, it has multi-dimensional irreducible representations along with one-dimensional ones, including the trivial representation.
Generally, we expect that the leading eigenvector of $T$ belongs to a one-dimensional irreducible representation, which implies that the strange correlator decays exponentially as in Eq.~\eqref{eq:Bij-l12}.
In fact, for a transfer matrix of a classical statistical system, its matrix elements must be all positive, and the leading eigenvector will belong to the trivial representation according to the Perron-Frobenius Theorem.
This is consistent with the fact that a 1D classical system cannot have spontaneous symmetry breaking.
However, the transfer matrix for a strange correlator generally has imaginary entries and therefore is not restricted by the above argument.
If the leading eigenvector belong to a nontrivial multi-dimensional (linear) irreducible representation of $G_b$, we would also expect a long-range correlation as in Eq.~\eqref{eq:Bij-c}.
However, in all examples we studied, the leading eigenvector always belong to a 1D irreducible representation.
Here, we conjecture that this is always true for transfer matrices generated by a trivial 2-coboundary, and leave a rigorous proof to future works.

Next, we generalize the above result to the case where $G_b$ contains antiunitary operations, which is labeled by a $\mathbb Z_2$ grading $s: G_b\rightarrow\mathbb Z_2$: $s(g)=0$ (1) indicates that $g$ is unitary (antiunitary), respectively.
Naturally, we should consider the corepresentations, instead of representations, of the symmetry group $G_b$.
(A brief review of the concept of corepresentation is included in Appendix~\ref{app:corep}.)
Since the cocycle condition of $\nu_2$ is modified by $s$, the above formulas need to be modified as the following:
First, the definition of $\tilde R(g)$ becomes the following:
\begin{equation}
    \label{eq:tRg-au-def}
    \tilde R(g)_{ij}=\nu_2(e, g, g_i)^{-1}R(g)_{ij}K^{s(g)},
\end{equation}
where $K$ denotes the complex-conjugate operation.
In fact, when $g$ is antiunitary, $\tilde R(g)$ is an antilinear operator on the Hilbert space instead of a linear operator (represented as a matrix).
Therefore, $\tilde R(g)$ forms a projective corepresentation of $G_b$:
\begin{equation}
    \label{eq:tRg-mul}
    \tilde R(g)K^{s(g)}\tilde R(h)K^{-s(g)} = \nu_2(e, g, gh)^{-1}\tilde R(gh).
\end{equation}
where $K$ denotes the complex-conjugate operation: $KxK^{-1}=x^\ast$.
Hence, $K^{s(g)}xK^{-s(g)}$ is $x$ ($x^ast$) if $g$ is unitary (antiunitary), respectively.
The intertwiner condition in Eq.~\eqref{eq:TR=RT} is also modified as the following,
\begin{equation}
    \label{eq:TR=RT}
    \tilde R(g) T =K^{s(g)}TK^{-s(g)}\tilde R(g).
\end{equation}
This allows us to apply Schur's lemma to obtain a block-diagonalized form of $T$.
However, the Schur's lemma has a more complicated form for corepresentations.
In fact, an intertwiner, such as $T$, is still zero between subspace of different irreducible corepresentations.
However, on the same subspace, $T$ can no longer be reduced to a scalar.
Instead, the possible choices of such intertwiner form a so-called division algebra over the field of real numbers.
Since there are three types of division algebras: the real numbers $\mathbb R$, the complex numbers $\mathbb C$ and the quaterions $\mathbb H$, the corepresentation can also be classified into three types:
a corepresentation is of type a, b, and c, if the division algebra of its intertwiners is $\mathbb R$, $\mathbb H$, and $\mathbb C$, respectively.
On the subspace of a type-a corepresentation, matrix $T$ will have degenerate real eigenvalues.
In contrast, on the subspace of a type-b or type-c corepresentation, the eigenvalues will not be degenerate: they will be complex and appear in complex-conjugate pairs as $\lambda$ and $\lambda^\ast$.

When $\nu_2$ is trivial, $\tilde R$ form a linear corepresentation of $G_b$, which can be decomposed into 1D type-a irreducible corepresentations (1D corepresentations must be type-a) and multi-dimensional ones.
Similar to the unitary case, we again assume that the leading eigenvector of $T$ belongs to a 1D type-a irreducible representation, and the strange correlator will decay exponentially as in Eq.~\eqref{eq:Bij-l12}.
When $\nu_2$ is nontrivial, $\tilde R$ form a projective corepresentation, which has only multi-dimensional irreducible corepresentations, which can belong to either one of the three types.
Hence, the leading vectors of $T$ will form a multi-dimensional corepresentation.
If the corepresentation belongs to type a, it will have degenerate real eigenvalues, and the strange correlator will tend to a constant as in Eq.~\eqref{eq:Bij-c}.
On the other hand, if the corepresentation belongs to type b or c, it will have two eigenvalues that are complex conjugate to each other.
As a result, the denominator and the numerator in Eq.~\eqref{eq:bs-corr-T} will have two equally dominating terms proportional to $\lambda^N$ and $\lambda^{\ast N}$, respectively.
We then expect the following behavior of the strange correlator,
\begin{equation}
    \label{eq:Bij-osc}
    C(i, j)\sim A+B\left[\left(\frac{\lambda^\ast}\lambda\right)^{|i-j|}+\text{c.c.}\right],
\end{equation}
where $A$ and $B$ are constants determined by the matrix elements of $O$ and $T$ on the irreducible corepresentation.
This means that the strange correlator can have an oscillationary behavior with a constant amplitute, when $\nu_2$ is nontrivial and the leading projective correpresentation belongs to type b or type c.

In summary, we see that for bosonic SPT states, the strange correlator of order parameters can detect whether the state is nontrivial, consistent with the general expectation~\cite{you_symmetry-protected_2014}: when $\nu_2$ is trivial, the strange correlator decays exponentionally as in \eqref{eq:Bij-l12}; when $\nu_2$ is nontrivial, the strange correlator is either a constant as in \eqref{eq:Bij-c}, or oscillates with a constant amplitude as in \eqref{eq:Bij-osc}.

\subsection{Fermionic strange correlator}
\label{sec:corr-f}


To detect the fermionic part FSPT phase, we use the complex-fermion annihilating operators $c_{i,i+1}$ as the local observables $O(r)$, and consider the following strange correlator,
\begin{equation}
    \label{eq:Cij}
    C(i,j)=\frac{\langle\Omega|c_{i,i+1}, c_{j,j+1}|\Psi\rangle}{\langle\Omega|\Psi \rangle},
\end{equation}
where $|\Psi\rangle$, the wave function being tested, has the form of Eq.~\eqref{eq:1Dwf}, and we choose $|\Phi\rangle$ to be a trivial FSPT wave function.

To compute this correlation function, we first expand the wave functions over the group-element-configuration basis,
\begin{equation}
    \label{eq:Cij-sum}
    \begin{split}
    &C(i, j)=\\
    &\frac{\sum_{\{g_i\}}
    \langle\{g_i\}|c_{i,i+1}c_{j,j+1}\nu_2(e, g_i, g_{i+1})|\{n_1(g_i,g_{i+1})\};\{g_i\}\rangle}
    {\sum_{\{g_i\}}
    \langle\{g_i\}|\nu_2(e, g_i, g_{i+1})|\{n_1(g_i,g_{i+1})\};\{g_i\}\rangle}
    \end{split}
\end{equation}
Because different Fock states are orthogonal,
the overlap between two states is nonvanishing only when the fermion decoration in the bra and ket states is identical.
Since the bra states $\langle\{g_i\}|$ has no fermion,
the overlap in the numerator of \eqref{eq:Cij} is nonvanishing if and only if
$|\{n_1(g_i,g_{i+1})\};\{g_i\}\rangle$ has exactly two fermions at bonds $(i,i+1)$ and $(j, j+1)$, and the overlap in the denominator of \eqref{eq:Cij} is nonvanishing when $|\{n_1(g_i,g_{i+1})\};\{g_i\}\rangle$ has no fermion.
Therefore, we need to sum over subset of group elements such that $n_1(g_i, g_{i+1})$ is 0 and 1, respectively.
In fact, when $n_1$ is a nontrival cocycle, $G_b$ can be divided into two cosets, such that $n_1(g_i, g_{i+1})=0$ if $g_i$ and $g_{i+1}$ belong to the same coset, and $n_1(g_i, g_{i+1})=1$ if otherwise.

We notice that $n_1\in H^1(G_b, \mathbb Z_2)$ can be interpreted as a group homomorphism $\rho: G_b\rightarrow\mathbb Z_2$, such that $n_1(g_i, g_{i+1})=\rho(g_i^{-1}g_{i+1})$, or a $\mathbb Z_2$ grading of $G_b$.
Therefore, the kernel of $\rho$ forms a normal subgroup of $G_b$, which we denote by $G_b^0=\rho^{-1}(0)$.
When $n_1$ is trivial, $\rho$ is also a trivial homomorphism that maps all elements to 0.
Therefore, $G_b^0=G_b$.
In this case, it is straightforward to see that $C(i, j)=0$, because there is no fermion in any configuration in $|\Psi\rangle$.
When $n_1$ is nontrivial, $\rho$ is surjective and $G_b/O=\mathbb Z_2$.
Therefore, $G_b$ can be divided into two cosets, one of which is $G_b^0$ and the other we denote by $G_b^1$.
$G_b^1$ can be viewed as $\sigma G_b^0$, where $\sigma$ is an arbitrarily chosen element in $G_b^1$.
Then, from the definition of $\rho$, it is straightforward to check that $n_1(g_i, g_{i+1})=0$ (or 1) if $g_i$ and $g_j$ belong to the same (different) cosets, respectively.
Using this $\mathbb Z_2$ grading, the nonvanishing entries in the summation in Eq.~\eqref{eq:Cij-sum} can be summarized as the following:
In the denominator, all group elements must belong to the same coset; in the numerator, group elements on vertices in the interval between $i$ and $j$ must belong to one coset, and elements on vertices outside this interval must belong to the other coset.

Following Sec.~\ref{sec:bosonic}, we can express the summation in Eq.~\eqref{eq:Cij-sum} using the transfer matrix $T_{ij}=\nu_2(e, g_i, g_j)$.
Since the summation involves the $\mathbb Z_2$ grading $G_b=G_b^0\cup G_b^1$, we divide $T$ into blocks: $T^{ab}_{ij}=\nu_2(e, \sigma^ag_i, \sigma^bg_j)$ (here $a,b=0,1$ denotes the grading, and $g_i,g_j\in G_b^0$).
Using this notation, Eq.~\eqref{eq:Cij-sum} is expressed as
\begin{widetext}
\begin{equation}
    \label{eq:Cij-trT}
    C(i, j)=\frac{\tr\left[T^{01}(T^{11})^{|i-j|}T^{10}(T^{00})^{N-|i-j|}\right]
    +\tr\left[T^{10}(T^{00})^{|i-j|}T^{01}(T^{11})^{N-|i-j|}\right]}{\tr\left[(T^{00})^N\right]+\tr\left[(T^{11})^N\right]}
\end{equation}
Next, we notice that the two terms in both the denominator and the numerator are related to each other by the symmetry operation $\sigma$.
To be precise, using the cocycle equation of $\nu_2$, it is straightforward to verify that 
\begin{equation}
    \label{eq:T11-T00}
    T^{11}=K^{s(\sigma)} S_\sigma T^{00}S_\sigma^{-1} K^{-s(\sigma});\quad
    T^{01}=K^{s(\sigma)}S_\sigma T^{10}S_\sigma^{-1}K^{-s(\sigma)},
\end{equation}   
\end{widetext}
where the matrix $S_\sigma$ is given by $(S_\sigma)_{ij}=\delta_{ij}\nu_2(e, \sigma, \sigma g_i)$.
In other words, $T^{11}$ ($T^{10}$) is similar to $T^{00}$ ($T^{01}$), respectively, if $\sigma$ is unitary,
and $T^{11}$ ($T^{10}$) is similar to the complex conjugate of $T^{00}$ ($T^{01}$), respectively, if $\sigma$ is antiunitary.
Hence, only one term in the denominator and numerator in Eq.~\eqref{eq:Cij-trT} needs to be computed, and the other term can be obtained by symmetry.

The argument of Sec.~\ref{sec:bosonic} suggests that the symmetry transformation of the transfer matrix plays an important role in determining its long-range behavior.
This motivates us to study the symmetry transformation of matrices $T^{ab}$.
In fact, we can generalize the intertwining condition \eqref{eq:TR=RT} to the following form,
\begin{equation}
    \label{eq:TR=RT-ab}
    \tilde R^a(g)T^{ab}= K^{s(g)}T^{ab}K^{-s(g)}\tilde R^b(g)
\end{equation}
for any $g\in G_b^0$.
Here, $\tilde R^a(g)$ denotes that the two vector spaces denoted by the grading $a,b=0,1$ form different representations of $G_b^0$ (or corepresentation if there are antiunitary symmetries; in this section, we can treat representation and corepresentation on equal footing, so we will generally refer to them as ``representation'' when it does not create confusion):
For $a=0$, the definition of $\tilde R^0(g)$ is similar to Eq.~\eqref{eq:tRg-def}
\begin{equation}
    \tilde R^0(g)_{ij} = R(g)_{ij}\nu_2(e, g, g_i)^{-1};
\end{equation}
for $a=1$, the definition is the following,
\begin{equation}
    \label{eq:tRg1-def}
    \tilde R^1(g)_{ij} = R(\sigma^{-1}g\sigma)_{ij}\nu_2(e, g, \sigma g_i)^{-1}.
\end{equation}
We notice that $g\mapsto \sigma^{-1}g\sigma$ is an inner automorphism of $G_b$, so $R(\sigma^{-1}g\sigma)_{ij}$ is actually the same as $R(g)_{ij}$.
In fact, the above two equations can be unified into the following definition,
\begin{equation}
    \label{eq:tRg01-def}
    \tilde R^a(g)_{ij} = R(g)_{ij}\nu_2(e, g, \sigma^ag_i)^{-1}.
\end{equation}
It is then straightforward to generaize Eq.~\eqref{eq:tRg-mul} and verify that the representations $\tilde R^a$ satisfies the same twisted relation,
\begin{equation}
    \label{eq:tRg01-mul}
    \tilde R^a(g)K^{s(g)}\tilde R^a(h)K^{-s(g)} = \nu_2(e, g, gh)^{-1}\tilde R^a(gh).
\end{equation}
Therefore, the two representations are equivalent projective representations.
In fact, they are related by the following gauge transformation,
\begin{equation}
    \label{eq:tRg01-gt}
    \tilde R^1(g)_{ij} = \frac{\nu_2(e, g, g_i)}{\nu_2(e, g, \sigma g_i)}\tilde R^0(g)_{ij}.
\end{equation}
This implies that they have identical decomposition to irreducible representations.
Moreover, $T^{ab}$, as intertwiner between $\tilde R^a$ and $\tilde R^b$, are block-diagonalized in each subspace of irreducible representations, as dictated by Schur's lemma. 

From the above symmetry analysis, we see that the eigenvectors of $T^{00}$ and $T^{11}$ carry the same irreducible representations.
Furthermore, according to Eq.~\eqref{eq:T11-T00}, the corresponding eigenvectors of the two matrices have identical (complex-conjugate) eigenvalues if $\sigma$ is unitary (antiunitary), respectively.
Hence, if we consider the leading eigenvector or eigenvectors dominating $(T^00)^{N-|i-j|}$, the matrix $T^{10}$ will map it to the corresponding eigenvector or eigenvectors of the same irreducible representation in $T^{11}$, which have the same (or complex-conjugate) eigenvalue.
This implies that the strange correlator $C(i, j)$ in Eq.~\eqref{eq:Cij} will always have the behaviors in Eq.~\eqref{eq:Bij-c} or Eq.~\eqref{eq:Bij-osc}.

In summary, we see that regardless whether $\nu_2$ is nontrivial or not, the strange correlator will generally have a long-range order when $n_1$ is nontrivial, and vanishes when $n_1$ is trivial.
Hence, it can be used to detect whether the fixed-point wave function has a nontrivial fermion decoration.

\subsection{Diagnose FSPT states}
\label{sec:diagnose}

We can combine the bosonic strange correlators and fermionic strange correlators studied in the previous sections to fully diagnose an FSPT state.

First, we use the fermionic strange correlator in Eq.~\eqref{eq:Cij} to detect if $|\Psi\rangle$ has a nontrivial $n_1$: the strange correlator will exhibit a long-range order if $n_1$ is nontrivial.
If $n_1$ is nontrivial, we need to further determine the cohomology class of $n_1$.
For this, we can use a different reference state $|\Phi\rangle$ with a nontrivial $n_1^0$.
In this case, the strange correlator will exhibit a long-range order if $n_1\neq n_1^0$, and an exponential decay if $n_1=n_1^0$.
Enumerating all possible cohomology classes in $H^1(G, \mathbb Z_2)$, we can determine the cohomology class of $n_1$.

Next, once $n_1$ is determined, we can use the bosonic strange correlator in Eq.~\eqref{eq:bs-corr} to further determine the bosonic phase factor $\nu_2$.
When $|\Psi\rangle$ has a nontrivial $n_1$, the bosonic strange correlator will have a more complicated form, because the transfer matrices are restricted to the blocks of $T^{00}$ and $T^{11}$.
In this way, we can only directly detect the restriction of $\nu_2$ on $G_b^0$.
To overcome this problem, we can use a different reference state $|\Phi\rangle$, with the same $n_1$ as $|\Psi\rangle$, which we have detected using the fermionic strange correlator.
In this way, for a given group-element configuration $\{g_i\}$, the basis states in $|\Phi\rangle$ and $|\Psi\rangle$ have the same fermion decoration and therefore are identical.
This allows the strange correlator to be expressed as Eq.~\eqref{eq:bs-corr-T}, and implies that it has the same behavior as discussed in Sec.~\ref{sec:bosonic}.
In this way, we can further determine $\nu_2$ using the bosonic strange correlator.

\section{1D FSPT states with $G_f=G_b\times_{\omega_2}\mathbb Z_2^f$}
\label{sec:gb-z2f-ext}

In this section, we explain how the results of previous section can be adapted to the case where the fermion symmetry group $G_f$ is a nontrivial extension of $G_b$ over $\mathbb Z_2^f$.

\subsection{Fixed-point wave function}
\label{sec:fpwf-ext}

First, the classification of FSPT states is modified~\cite{wang_construction_2020}:
choices of $n_1$ is now a subset of $H^1(G_b, \mathbb Z_2)$ such that the obstruction
\begin{equation}
    \label{eq:O3}
    O_3[n_1] = (-1)^{\omega_2\cup n_1}
\end{equation}
belongs to the trivial cohomology class in $H^3[G_b, \mathrm U(1)]$, and $\nu_2$ is no longer a cocycle but satisfies the following twisted cocycle equation,
\begin{equation}
    \label{eq:dnu2}
    d\nu_2 = (-1)^{\omega_2\cup n_1}.
\end{equation}
Furthermore, $\nu_2$ that differs by $(-1)^{\omega_2}$ are considered identical FSPT states.

Correspondingly, the construction of fixed-point wave functions needs to be modified.
First, since the fermions now transform projectively under $G_b$, then cannot be presented by spinless fermions.
Instead, the fermions carry a flavor index $g\in G_b$, and transforms as
\begin{equation}
    \label{eq:gc-w}
    g\cdot c^h=(-1)^{\omega_2(e, g, gh)}c^{gh}.
\end{equation}
The fermion decoration is determined by $n_1$ in a similar way as in Sec.~\ref{sec:wavefunction}, but we also need to specify which flavor to decorate.
On the bond $\langle i,i+1\rangle$, if $n_1(g_i, g_{i+1})=1$, we decorate $c^{g_i}_{i, i+1}$ on this bond, which is expressed as $n_{i,i+1}^g = \delta_{g, g_i}n_1(g_i, g_{i+1})$.
Second, the wave function has a bosonic phase factor given by the cochain $\nu_2$.
Overall, the wave function can be expressed as
\begin{equation}
    \label{eq:Psi-w2}
    |\Psi\rangle=\sum_{\{g_i\}}
    \prod_i\nu_2(e, g_i, g_{i+1})\prod_i\left(c_{i,i+1}^{g_i\dagger}\right)^{n_1(g_i, g_i+1)}|\{g_i\}\rangle.
\end{equation}


\subsection{Fermionic strange correlator}
\label{sec:corr-f-ext}

We now generalize the discussion of fermionic strange correlator in Sec.~\ref{sec:corr-f} to the cases of nontrivial $\omega_2$.
We still want to use the complex-fermion annihilation operator as the observables.
However, since the fermion operator now carries an additional flavor, we need to decide which flavor, or what kind of combination of flavors to include in the observable.
Ideally, we want to construct observable operator $\hat O_i$ that is invariant under $G_b$ action, as the operator $c$ in Sec.~\ref{sec:corr-f}.
Unfortunatelly, this appears to be impossible because the fermion annihilation operator now carries projective representation of $G_b$, and therefore their linear combination cannot carry the trivial linear representation.
In the following, we describe two possible strategies to deal with this difficulty, which both works in most cases but are both less than ideal.

Generally, we can define a linear superposition of different flavors of fermion operators,
\begin{equation}
    \label{eq:c-phi}
    c = \sum_{g\in G_b} \phi(g)c_g,
\end{equation}
and use it to construct a strange correlator.
Similar to the discussion in Sec.~\ref{sec:corr-f}, we obtain a modified version of Eq.~\eqref{eq:Cij-trT},
\begin{widetext}
\begin{equation}
    \label{eq:Cij-trT}
    C(i, j)=\frac{\tr\left[T^{\prime 01}(T^{11})^{|i-j|}T^{\prime 10}(T^{00})^{N-|i-j|}\right]
    +\tr\left[T^{\prime 10}(T^{00})^{|i-j|}T^{\prime 01}(T^{11})^{N-|i-j|}\right]}{\tr\left[(T^{00})^N\right]+\tr\left[(T^{11})^N\right]},
    \end{equation}    
\end{widetext}
where the modified transfer matrices $T^{\prime 10}$ and $T^{\prime 01}$ are given by
\begin{equation}
    \label{eq:Tprime-phi}
    T^{\prime ab}_{ij} = \nu_2(e, \sigma^ag_i, \sigma^bg_j)\phi(g_i).
\end{equation}
Here, it can be verified that $T^{00}$ and $T^{11}$ still satisfies the intertwining condition in Eqs.~\eqref{eq:TR=RT-ab}.
In fact, although the proof in Sec.~\ref{sec:corr-f} uses the cocycle condition of $\nu_2$, which is violated here as in Eq.~\eqref{eq:dnu2},
the violation involves $d\nu_2(e, g, g_i, g_j)$ or $d\nu_2(e, g, \sigma g_i, \sigma g_j)$, which actually vanishes as the last two group elements belong to the same coset and therefore $n_1(g_i, g_j)=n_1(\sigma g_i, \sigma g_j)=0$.
Therefore, the eigenvectors of $T^{00}$ and $T^{11}$ still forms irreducible representations as in $tilde R^{0,1}$, respectively.
However, $T^{\prime 10}$ and $T^{\prime 01}$ no longer satisfies the intertwining condition between $\tilde R^{0,1}$.
In fact, take $T^{\prime 10}$ as an example, we get the following relation instead,
\begin{equation}
    \label{eq:TR=RT-w}
    \tilde R^1(g)_{ij}T^{\prime 10}_{jk}
    = (-1)^{\omega_2(e, g, \sigma g_i)}K^{s(g)}\frac{\phi(\sigma g_i)K^{-s(g)}}{\phi(g_i)}T^{\prime 10}_{ij}\tilde R^0(g)_{jk}
\end{equation}
Here, additional phase factors appear in the intertwining condition.
In particular, the 2-cocycle $\omega_2$ appears because of Eq.~\eqref{eq:dnu2}.
These phase factors mean that $T^{\prime10}$ no longer maps the eigenvectors of $T^{00}$ to the corresponding ones in $T^{11}$.

To deal with this problem, we propose two approaches:
First, we can simply discard the symmetry of $G_b^0$, and use an arbitrary $\phi(g)$ which may not even form a representation of $G_b$.
For example, we may simply choose $\phi(g)=\delta_{g,e}$ or $\phi(g)=1$.
In this case, since the symmetry has been explicitly broken, the overlap $\langle \alpha^1|T^{\prime 10}|\alpha^0\rangle$, where $|\alpha^{0,1}\rangle$ is the leading eigenvector of $T^{00}$ and $T^{11}$, respectively, should generally be nonzero without fine tuning.

Second, we can assemble an operator $c$ that forms a projective representation of $G_b$.
When $(-1)^{\omega_2}$ is a trivial cohomology class in $H^2[G_b, \mathrm U(1)]$~\footnote{A nontrivial cocycle $\omega_2\in H^2(G_b, \mathbb Z_2)$ can induce a trivial cocycle $(-1)^{\omega_2}$, as more coboundaries with arbitrary values in $\mathrm U(1)$ instead of only $\pm1$ are included.}, such projective representation can be constructed as one-dimensional.
Otherwise, it must be multi-dimensional and the resulting operator will carry an additional flavor index.
For simplicity, we only consider the former case, and leave the generalization to the latter case to future works.
We assume $(-1)^{\omega_2}$ can be expressed as a coboundary:
\begin{equation}
    \label{eq:w2=da}
    \omega_2(g_1, g_2, g_3) = \frac{\alpha(g_2, g_3)\alpha(g_1, g_2)}{\alpha(g_1, g_3)},
\end{equation}
and choose $\phi(g)=\alpha(e, g)$.
The operator constructed as $c=\sum_g\phi(g)c^g$ then transforms as a one-dimensional projective representation $g\cdot c=\phi(g)c$.
In other words, the operator $c$ carries a 1D projective representation of $G_b$ [1D representations exist because $(-1)^{\omega_2}$ is trivial].
The phase factor in Eq.~\eqref{eq:TR=RT-w} is also simplified as follows,
\begin{equation}
    \label{eq:TR=RT-w}
    \tilde R^1(g)_{ij}T^{\prime 10}_{jk}
    = \phi(g)K^{s(g)}T^{\prime 10}_{ij}K^{-s(g)}\tilde R^0(g)_{jk}
\end{equation}
Hence, the matrix $T^{\prime 10}$ maps representations in $\tilde R^0$ to their tensor product with the (1D) projective representation $\phi(g)$.
Since when $(-1)^{\omega_2}$ is trivial, $\mathbb C[G_b]^{(-1)^{\omega_2}}$ is isomorphic to $\mathbb C[G_b]$, the projective representation with factors $(-1)^{\omega_2}$ also forms a complete orthogonal basis of the Hilbert space spanned by $|g\rangle$.
Therefore, in general, the overlap between basis vectors in two sets of basis is nonzero.
This implies that $\langle \alpha^1|T^{\prime 10}|\alpha^0\rangle\neq0$ in general, without fine tuning.

In summary, we observe that because the fermion operators transform projectively under $G_b$, it is impossible to construct local fermionic operators that trasform trivially under $G_b$ (in fact, the bosonic degrees of freedom transform linearly, and therefore cannot cancel the projective representation).
As a result, the corresponding transfer matrix will not map the leading eigenvector of $T^{00}$ to only the corresponding eigenvector in $T^{11}$.
However, the transfer matrix still has a nonvanishing matrix element between the two leading eigenvectors, $\langle \alpha^1|T^{\prime 10}|\alpha^0\rangle\neq0$.
This still implies that the strange correlator goes to a constant or oscillates with a constant amplitude, as in Eqs.~\eqref{eq:Bij-c} and \eqref{eq:Bij-osc}.


\subsection{Diagnose FSPT states}
\label{sec:diagnose-ext}

Since the fermionic strange correlator still works for the cases with a nontrivial $\omega_2$, we can still use it, together with bosonic strange correlators, to fully diagnose an FSPT state, using a procedure similar to Sec.~\ref{sec:diagnose}.

First, we determine $n_1$ using the fermionic strange correlator, by trying possible obstruction-free solutions of $n_1$ in the reference state.
Next, we construct a reference state with previously determined $n_1$, and a special solution $\nu_2^0$ of the twisted cocycle equation \eqref{eq:dnu2}.
Then, the difference between $\nu_2$ of $|\Psi\rangle$ and $\nu_2^0$ is a (untwisted) cocycle in $H^2[G_b,\mathrm U(1)]$.
This difference is then detected using bosonic strange correlators discussed in Sec.~\ref{sec:bosonic}.
Following this procedure, we can fully diagnose an FSPT using a combination of strange correlators.

\section{Examples}
\label{sec:examples}
\subsection{$G_f=\mathbb Z_2^T\times\mathbb Z_2^f$}
We start with the simple example of $G_f=\mathbb Z_2^T\times\mathbb Z_2^f$.
It is easy to compute that $H^2[\mathbb Z_2^T,\mathrm U(1)]=\mathbb Z_2$ and $H^1(\mathbb Z_2^T,\mathbb Z_2)=\mathbb Z_2$, so the classification of $n_1$ and $\nu_2$ is both $\mathbb Z_2$.
In fact, this is known as the BDI class in the ten-fold way, and the classification of topological phases for interacting fermions in this symmetry class is known to be $\mathbb Z_8$~\cite{Fidkowski2010, fidkowski_topological_2011}.
The root phase of the $\mathbb Z_8$ classification is a Kitaev chain (also known as a Majorana chain), which is not considered an SPT state because it does not need the protection of the time-reversal symmetry in this work.
Also, there is a nontrivial extension between the two layers~\cite{ren2023stacking}: stacking two identical states with nontrivial $n_1$ produces a state with a nontrivial $\nu_2$.
This explains how the $\mathbb Z_8$ classification is obtained using the approach in Ref.~\onlinecite{wang_towards_2018}.

In the following, we consider the nontrivial FSPT state $|\Psi\rangle$ with both a nontrivial $n_1$ and a nontrivial $\nu_2$.
The nontrivial $n_1$ is given by $n_1(e, T)=0$ and $n_1(e, T)=1$, and the nontrivial $\nu_2$ is given by $\nu_2(e, e, e)=\nu_2(e, e, T)=\nu_2(e, T, T)=1$ and $\nu_2(e, T, e)=-1$. (All other entries can be obtained by symmetry conditions of the homogeneous cocycle.)

To detect $n_1$, we select $|\Phi\rangle$ to be the trivial state with $n_1=0$ and $\nu_2=1$.
With these choices, we can proceed to compute the strange correlator:
\begin{align}
\label{eq:1Dwf}
C(i,j)=\frac{\langle\Omega|c_{i,i+1}, c_{j,j+1}|\Psi\rangle}{\langle\Omega|\Psi \rangle}=-1
\end{align}
This result indicates that $n_1$ is nontrivial.

Next, to detect the bosonic part $\nu_2$, we select a non-trivial reference state where the 1-cocycle $n_1$ is the same as that of the state under measurement
We consider the order parameters $O_i$ and $O_j$ as our local observables (the order parameter is $\pm1$ if $g_i$ is $T$ and $e$, respectively). With these choices, we can compute the strange correlator as follows:
\begin{equation}
\label{eq:1Dwf}
C(i,j)=\frac{\langle\Omega|O_iO_j|\Psi\rangle}{\langle\Omega|\Psi \rangle}=\begin{cases}
1\ \ (|i-j|=0, 1 \mod 4)
\\
-1\ \ (|i-j| = 2, 3\mod\ 4)
\end{cases}
\end{equation}
This result demonstrates oscillation, which is consistent with the predictions of the strange correlation theory.
In fact, for the nontrivial $\nu_2$, the eigenvalues of the transfer matrix $T$ defined in Sec.~\ref{sec:bosonic} are $1\pm i$, indicating a two-fold degenerate projective corepresentation of type b.
This observation explains why the strange correlator of the non-trivial SPT state of $G_b = Z_2^T$ exhibits oscillations as in Eq.~\eqref{eq:Bij-osc}.


\subsection{$G_f=\mathbb Z_4\times\mathbb Z_4\times\mathbb Z_2^f$}
\label{sec:z4z4z2f}

Next, we consider the example of $G_f=\mathbb Z_4\times\mathbb Z_4\times\mathbb Z_2^f$.
The classification of fermion decoration $n_1$ is given by $H^1(\mathbb Z_4\times\mathbb Z_4, \mathbb Z_2) =\mathbb Z_2\times\mathbb Z_2$, and the classification of bosonic phase factor $\nu_2$ is given by $H^2[\mathbb Z_4\times\mathbb Z_4, \mathrm U(1)]=\mathbb Z_4$.
In other words, there are three nontrivial choices of $n_1$.
In fact, the three choices of $n_1$ are equivalent to each other up to an automorphism of $G_b$.
Therefore, without losing generality, we can focus on one of them.
There are three nontrivial choices of $\nu_2$, which can be expressed as a generator cocycle $\alpha_2$, and its second and third powers $\alpha_2^2$, and $\alpha_2^3$, respectively.
Here, $\nu_2^m$ denotes the cocycle whose entries are $\nu_2^m(g_0, g_1, g_2)=[\nu_2(g_0,g_1,g_2)]^m$.

We denote an element of $G_b=\mathbb Z_4\times\mathbb Z_4$ as $a^mb^n$, where $a$, $b$ are generators of the first and second $\mathbb Z_4$ in $G_b$, respectively, and $m, n=0,1,2,3$.
In this way, the nontrivial $n_1$ we choose to consider is $a_1(e, a^mb^n)=m \mod 2$, and the generator cocycle $\nu_2$ is expressed as
$\alpha_2(e, a^mb^n, a^{m'}b^{n'})=i^{m(n'-n)}$.

First, we study the fermion strange correlator.
For this, we assume the reference state is the trivial state with $n_1=0$ and $\nu_2=0$, and let the state being tested has $n_1=a_1$.
According to the notation in Sec.~\ref{sec:corr-f}, this implies that $G_b^0=\mathbb Z_2\times\mathbb Z_4$, generated by $a^2$ and $b$.
If the state being tested also has a trivial $\nu_2=1$, it is straightforward to check that the fermion strange correlator in Eq.~\eqref{eq:Cij-trT} is always 1 regardless of the distance $|i-j|$.
The situation is more complex if the state being tested has a nontrivial $\nu_2=\alpha_2$.
In this case, the leading eigenvectors of the transfer matrix $T^{00}$ is two-fold degenerate and form a projective representation of $G_b^0$, because the cocycle $\alpha_2$ is still nontrivial when restricted to $G_b^0$.
As shown in Sec.~\ref{sec:corr-f}, the matrix $T^{10}$ is a constant map between the two-fold-degenerate leading eigenvectors of $T^{00}$ and the corresponding two-fold-degenerate eigenvectors in $T^{11}$, but the constant is exactly zero.
This indicates that the strange correlator actually decays exponentially.
This can be understood as an accidental cancellation instead of a violation of general behaviors of strange correlators,
because such vanishing of the constant map in $T^{10}$ and the corresponding exponential decay will disappear if we perturb the cocycle $\nu_2$ away from the cochain $\alpha_2$ by an arbitrarily small coboundary.
For example, if we deform $\nu_2$ as $\nu_2=\alpha_2+d\beta_1$, where $\beta_1(e, a)$ is $e^{i\theta}$ and $\beta_1(e, g)=1$ for any other group elements, we shall see that the constant map in $T^{10}$ never vanishes except for $\theta=0,\pi$, and therefore the strange correlator always goes to a constant value at long distances, except at $\theta=0,\pi$ the constant value vanishes due to an accidental cancellation.
In practise, we usually do not need to worry about such accidental cancellations because it won't happen to a realistic ground-state wave function without fine tuning.
We leave the understanding of the reason behind this accidental cancellation to future studies.


Next, to detect  the bosonic part of 1D FSPT state, we use a reference state with 1-cocycle $n_1$ same as that of the state under measure and $\nu_2 = 1$.
As discussed in Sec.~\ref{sec:diagnose}, the bosonic strange correlator for this pair of states is then the same as a bosonic strange correlator between two bosonic SPT states.
We observe that the transfer matrix of the non-trivial cocycle of order 2 possesses two degenerate leading eigenvalues: $\lambda_1 = \lambda_2= 8$, if the test state has $\nu_2=\alpha^2$. Similarly, the transfer matrix of $\nu_2=\alpha$ or $\nu_2=\alpha^3$ possesses four degenerate leading eigenvalues: $\lambda_1=\lambda_2=\lambda_3=\lambda_4 = 4$.
As discussed in Sec.~\ref{sec:bosonic}, these degeneracies indicates that strange correlators decay to a constant in long distance, reflecting the nontrivial $\nu_2$ in the test state.

\subsection{$G_f=\mathbb Z_{2N_0}^f\times\mathbb \prod_{i=1}^K\mathbb Z_{N_i}$}
We now consider the case where the symmetry group $G_f$ is a central extension of the bosonic unitary finite Abelian group $G_b=Z_{N_0}\times\prod_{i=1}^K Z_{N_i}$ \cite{zhou_non-abelian_2021, wang_construction_2020, thomas_farrell_second_1974}, with $\omega_2(e,a,ab)=\left\lfloor\frac{a^0+b^0} {N_0}\right\rfloor$, where $\left\lfloor x\right\rfloor$ is the floor operator that picks out the largest integer less than or equal to $x$.
This means that the group structure of $G_f$ is $\mathbb Z_{2N_0}^f\times\mathbb \prod_{i=1}^K\mathbb Z_{N_i}$.
Furthermore, we focus on cases where $N_i=2^{k_i}$ because these are the ones that interesting FSPT phases are present, and assume without losing generality that $N_i\leq N_{i+1}$.



We denote an element $g\in G_b$ as a tuple of integer numbers $g=(g^0,g^1,g^2,\dots,g^K)$, where each component $g^i$ indicates that the projective of $g$ in the $i$-th subgroup $\mathbb Z_{N_i}$ is the generator of that group raised to the $g^i$-th power.
We then introduce the notation of the following 1-cocycles $n_1^i$, $i=1,\ldots,K$, whose entries are given by $n_1^i(e, g)=g^i\mod2$.
It is worth noting that $n_2^{0}$ is the nontrivial $Z_2$-valued 2-cocycle of $Z_{N_0}$, rather than that of $Z_{2N_0}$.

In the following, we focus on a test state with $n_1=n_1^i$. 
Using the ingredients above, we can solve the twisted cocycle equation
$d\nu_2=(-1)^{\omega_2\cup n_1}$.
We can prove that the right-hand side is a trivial cocycle if and only if $N_i>N_0$ or equivalently $k_i>k_0$.
Assuming so, it can then be verified that the following $\nu_2(g_0,g_1,g_2)$ satisfies the twisted cocycle equation:
\begin{align}
\nu_2(g_0,g_1,g_2)=e^{i\frac{\pi}{N_0}n_1^0(g_0,g_1)n_1^i(g_1,g_2)}.
\end{align}

As mentioned previously in Sec.~\ref{sec:corr-f-ext}, when $G_f$ is a nontrivial central extension of $G_b$, it is impossible to construct local fermion operator that is invariant under the symmetry actions.
Here, we select the complex fermion annihilating operators $c_{i,i+1}=\sum_{g_i\in G_b}c_{i,i+1}^{g_i}$ and $c_{j,j+1}=\sum_{g_j\in G_b}c_{j,j+1}^{g_j}$ as local observables in the system and use their strange correlator to detect the non-triviality of the system.
After performing a detailed and intricate calculation, we obtain the following expression for the strange correlator:
\begin{equation}
\label{eq:1Dwf}
\begin{split}
C(i,j)&=\frac{\langle\Omega|c_{i,i+1}, c_{j,j+1}|\Psi\rangle}{\langle\Omega|\Psi \rangle}\\
&=\frac{\sum_{m=0}^2 \binom{2}{m} e^{i(m\pi/N_0)}}{2^2}=\text{Const},    
\end{split}
\end{equation}
where $\binom{N}{m}$ denotes the combination coefficient.
This example demonstrates that when using complex fermion annihilation operators as detectors, the strange correlator of the 1D FSPT state is a constant, even when such operators are not invariant under symmetry.
As argued in Sec.~\ref{sec:corr-f-ext}, such a constant is not forbidden by symmetry and therefore generally does not vanish without fine tuning.

\section{Conclusions}
\label{sec:conclusion}

In this work, we study the strange correlator in 1D fermionic symmetry-protected topological (FSPT) states.
We confirm that the properties of the strange correlator remain valid for generic 1D FSPT phases.
Although we focus on fixed-point wave functions, our conclusion should qualitatively apply to generic FSPT wave functions with nonzero correlation lengths, because such a generic FSPT wave function become a fixed-point wave function of the same FSPT phase after real-space coarse graining, which does not affect the long-range behavior of strange correlators.
Furthermore, we provided specific examples that demonstrate the existence of both constant and oscillating strange correlators.

We note that our study only focused on the 1D case, and the strange correlator for FSPT states still require further investigation, especially in higher-dimensional systems which possess more degrees of freedom, such as the Kitaev Majorana chain and $p+ip$ superconductor decorations.
In particular, in our study of 1D strange correlators, the fact that the eigenvectors of transfer matrix of the corresponding statistical problem forms multi-dimensional projective representations plays an important role.
It will be interesting to see how this can be generalized to higher dimensions, where the symmetry anomaly is manifested not only as projective representations but also as nonassociative string operators, or as nontrivial entanglement between actions of $G_b$ and fermionic degrees of freedom.
We will leave these to future studies.

\begin{acknowledgments}
    Y.N. acknowledges Qing-Rui Wang, Tian Yuan, Fangji Liu, and Jian-Kang Huang for their insightful comments and helpful discussions.
    Y.Q. is supported by National Key R\&D Program of China (Grant No. 2022YFA1403400) and by NSFC Grant No. 12174068.
\end{acknowledgments}

\appendix
\section{Corepresentation of antiunitary symmetry groups}
\label{app:corep}

In this appendix, we briefly review the concept of corepresentations of symmetry groups containing antiunitary symmetry operations.
Mathematically, we present the concept of a symmetry group with antiunitary operations with a group $G$ and a $\mathbb Z_2$ grading $s$ specifying which elements are antiunitary.
$s: G\rightarrow \mathbb Z_2$ is a group homomorphism, and $s(g)=0$ (1) indicates that $g$ is unitary (antiunitary).

The concept of a corepresentation describes how such a symmetry group acts on a Hilbert space of quantum states.
To motivate this concept, consider an antiunitary element $g$.
In quantum mechanics, the action of $g$ on quantum states should be antilinear instead of linear:
\begin{equation}
    \label{eq:au-al}
    g(x|\alpha\rangle)=x^\ast g|\alpha\rangle,
\end{equation}
where $x^\ast$ denotes the complex conjugate of $x\in\mathbb C$.
In particular, since the action of $g$ does not commute with multiplying a complex number, it is no longer a linear map and not an element of $GL(V)$, where $V$ denotes the Hilbert space.
To better describe it, we can introduce a complex-conjugate operation (known as a real structure in mathematics) $K$ on the Hilbert space $V$.
The action of $g$ can then be expressed as an element of $GL(V)$ multiplies $K$.
In general, an action of $g\in G$ can be expressed as $\phi(g)K^{s(g)}$, where $\phi(g)$ is a linear map in $GL(V)$.
Therefore, we define a corepresentation of an antiunitary symmetry group $(G, s)$
as a vector space $V$, and a map $\phi:G\rightarrow GL(V)$, which is not a group homomorphism but satisfies the following condition,
\begin{equation}
    \label{eq:corep-phigh}
    \phi(g)K^{s(g)}\phi(h)K^{-s(g)}=\phi(gh).
\end{equation}

Next, we introduce the Schur's lemma for corepresentations.
We define an intertwiner $A$ (or an intertwining map) between two irreducible corepresentations $U$ and $V$ as a linear map $A:U\rightarrow V$ that commutes with the actions of $G$: $\phi_V(g)K^{s(g)}A=A\phi_U(g)K^{s(g)}$, or equivalently,
\begin{equation}
    \label{eq:intertwin}
    \phi_V(g)K^{s(g)}AK^{-s(g)}=A\phi_U(g).
\end{equation}
Similar to the unitary cases, Schur's lemma asserts that an intertwiner between two inequivalent irreducible corepresentations must be zero, and we only need to consider intertwiners between equivalent corepresentations.
For convenience, let us consider an intertwiner from a corepresentation $V$ to itself.
Schur's lemma asserts that, if $A$ is Hermitian, then it must be a scalar matrix.
This conclusion, when applied to the Hamiltonian of a quantum-mechanical system, explains the Kramer's degeneracy due to time-reversal symmetries.
However, $A$ may not be a scalar matrix if it is not Hermitian.
In fact, it can be proven that, on a irreducible corepresentation, possible choices of intertwiner $A$ form a real division algebra.
(A division algebra is an algebra in which all nonzero elements are invertible.)
Since there are three types of real division algebra: the real numbers $\mathbb R$, the complex numbers $\mathbb C$ and the quartonians $\mathbb H$, the irreducible corepresentations can also be classified into three types.
They are called type a, b, and c, if their intertwiners form the algebra of $\mathbb R$, $\mathbb H$ and $\mathbb C$, respectively.

\bibliography{reference,ref2}
\end{document}